\documentclass[referee]{raa}
\usepackage{graphicx,times}
\usepackage{natbib}
\usepackage{amssymb,amsmath}
\usepackage[section]{placeins}
\bibpunct{(}{)}{;}{a}{}{,}

\usepackage[pagebackref=true]{hyperref}

\begin{document}

   \title{Constraining the spatial curvature of the local Universe with deep learning}

 \volnopage{ {\bf 20XX} Vol.\ {\bf X} No. {\bf XX}, 000--000}
   \setcounter{page}{1}

   \author{Liang Liu \inst{1,2}, Li-Juan Hu\inst{1}, Li Tang\inst{1,2}, Ying Wu\inst{1,2}
   }

   \institute{ Department of Math and Physics, Mianyang Teachers' College, Mianyang 621000, China; {\it tang@mnu.edu.cn}\\
        \and
             Research Center of Computational Physics, Mianyang Teachers' College, Mianyang 621000, China\\
\vs \no
   {\small Received 20XX Month Day; accepted 20XX Month Day}
}

\abstract{We use the distance sum rule (DSR) method to constrain the spatial curvature of the Universe with a large sample of 161 strong gravitational lensing (SGL) systems, whose distances are calibrated from the Pantheon compilation of type Ia supernovae (SNe Ia) using deep learning. To investigate the possible influence of mass model of the lens galaxy on constraining the curvature parameter $\Omega_k$, we consider three different lens models. Results show that a flat Universe is supported in the singular isothermal sphere (SIS) model with the parameter $\Omega_k=0.049^{+0.147}_{-0.125}$. While in the power-law (PL) model, a closed Universe is preferred at $\sim 3\sigma$ confidence level, with the parameter $\Omega_k=-0.245^{+0.075}_{-0.071}$. In extended power-law (EPL) model, the 95$\%$ confidence level upper limit of $\Omega_k$ is $<0.011$. As for the parameters of the lens models, constrains on the three models indicate that the mass profile of the lens galaxy could not be simply described by the standard SIS model.
}

   \authorrunning{L. Liu et al. }            
   \titlerunning{Constraining $\Omega_k$ with DL}  
   \maketitle

%
\section{Introduction}           
\label{sect:intro}

The standard cosmological model ($\Lambda$CDM model), regarded as a cornerstone solution derived from the homogeneous and isotropic Friedmann-Robertson-Walker (FRW) metric, presents a comprehensive framework in cosmology. This model postulates the existence of radiation, ordinary baryonic matter, non-luminous dark matter, and enigmatic dark energy as constituents of the Universe. Its validity and credibility find strong support from a plethora of cosmological observations \citep{PlanckCollaboration:2016,PlanckCollaboration:2020}. Especially, the most recent findings derived from the conclusive full-mission analysis of the cosmic microwave background (CMB) anisotropies by the Planck mission exhibit remarkable agreement with the prevailing spatially-flat 6-parameter $\Lambda$CDM cosmological model. These results not only validate the standard framework but also provide stringent constraints on the cosmological parameters with an exceptional level of precision \citep{PlanckCollaboration:2020}. However, some challenges come following the success of $\Lambda$CDM model. Recently, the ``\,$H_0$ tension problem", i.e. the measured value of Hubble constant $H_0$ from the local Type Ia supernovae (SNe Ia) observation is inconsistent with the result from the Planck observation of CMB, has attracted great attention \citep{Freedman:2017,Riess:2016jrr,Riess:2019,DiValentinoaE.:2021}. This discrepancy is possibly caused by either unknown systematic uncertainties or new physics beyond the standard $\Lambda$CDM cosmology.

In the context of the FRW metric, the spatial curvature parameter plays a pivotal role in elucidating the geometric nature of the Universe. By incorporating measurements from the CMB and baryon acoustic oscillation (BAO), it has been established that the Universe can be reasonably modeled as spatially flat. This conclusion is supported by the constraint on the curvature parameter, $\Omega_k=0.001\pm0.002$ \citep{PlanckCollaboration:2020}. Considering the intricate degeneracy between the curvature parameter and the equation of state of dark energy, the assumption of a flat universe is commonly adopted during the analysis of dark energy properties. Small deviation of spatial curvature from zero would generate enormous effects on the reconstruction of dark energy and on the evolution of the Universe \citep{IchikawaandTakahashi:2006,Clarkson:2007,GongandWang:2007,Virey:2008}. Although the Planck CMB data constrains the spatial curvature at a very high precision, it not only depends on a certain cosmological model (the $\Lambda$CDM model), but also on the evolution of the early Universe. Recently, the reanalysis of Planck data showed that a closed universe is favoured against a flat universe \citep{DiValentinoaE.:2020,DiValentinoaE.:2021}. The presence of the so-called "$H_0$ tension problem" suggests the possibility of deviations between the actual state of the Universe and the predictions of the standard $\Lambda$CDM model. Specifically, it implies that the cosmological parameters derived from CMB measurements may differ from those obtained through local data. Consequently, it becomes crucial to ascertain the spatial curvature of the local Universe in a manner that is independent of specific theoretical models.

The measurement of spatial curvature is generally the by-product of the validity test of FRW metric. A model-independent approach was introduced by \cite{Clarkson:2007,Clarkson:2008} to scrutinize the validity of FRW metric. This method involves a comparative analysis of the cosmic expansion rate and cosmological distance, and has since been widely employed to examine the FRW metric and impose constraints on the spatial curvature \citep{MortsellandJonsson:2011, Sapone:2014, Cai:2016}. \cite{Bernstein:2006} put forth an alternative model-independent geometric approach to constrain spatial curvature. This methodology revolves around the fundamental sum rule of distances along null geodesics within the FRW metric framework. \cite{Rasanen:2015} employed the distance sum rule (DSR) to evaluate the accuracy of the FRW metric. By combining data from SNe Ia and strong gravitational lensing (SGL), they examined the validity of the FRW metric. Their analysis confirmed the overall validity of the FRW metric, although the obtained constraint on the spatial curvature parameter was relatively weak or loosely constrained. Considering the interdependence between the curvature parameter and the parameters of the lensing model, \cite{Xia:2017} adopted more intricate lensing models in their analysis and attained constraints on the spatial curvature by leveraging a substantial dataset comprising 118 SGL systems \citep{Cao:2015, Cao:2016}. Following this line, there are a series of works devoting to constraining the spatial curvature with updated observational data \citep{Li:2018,Qi:2018,Liu:2020a,Cao:2021}. It should be noted that the constraints on the spatial curvature derived from the aforementioned studies suffer from limitations arising from the relatively small size of the available SGL sample, as well as uncertainties stemming from unknown systematic effects. Moreover, the methods to calibrate the distances of lenses and sources within SGL systems rely on a polynomial approximation that is assumed to fit the SNe Ia sample. Further research and advancements in data acquisition and analysis techniques are necessary to address these limitations and improve the precision of spatial curvature measurements in cosmology.

To alleviate the above shortcomings, \cite{Wang:2020} made significant advancements in constraining the spatial curvature. They employed the DSR method and combined data from the Pantheon SNe Ia compilation with a dataset comprising 161 SGL systems. Notably, they avoided assumptions regarding the parametric form of the distance-redshift relation of SNe Ia. Instead, they employed a Gaussian Process (GP) method to reconstruct the dimensionless comoving distance based on the Pantheon compilation. Without the prior of $H_0$, the constraints on spatial curvature are $\Omega_k=0.57^{+0.20}_{-0.28}$ in the singular isothermal sphere model, $\Omega_k=-0.246^{+0.078}_{-0.100}$ in the power-law model, and $\Omega_k=0.25^{+0.16}_{-0.23}$ in the extended power-law model. Previous studies have indicated that a larger dataset is beneficial to achieve a tighter constraint on $\Omega_k$ \citep{Xia:2017,Qi:2018,Li:2018}. However, the GP method is unable to extrapolate the curve beyond the available data region, and its accuracy diminishes significantly in regions where data points are sparse. Consequently, in their analysis, only the SGL systems with redshifts lower than the maximum redshift of the SNe Ia data could be utilized. This constraint resulted in a reduction in the number of available SGL systems from the initial 161 to 135. Therefore, although the constraint of \cite{Wang:2020} is tighter than previous works, the method to reconstruct the distance-redshift relation can be further improved so that all SGL systems can be used to constrain the spatial curvature.

In this paper, we will maintain the advantages of \cite{Wang:2020}, i.e. the large dataset and model-independence, and employ a deep learning method to reconstruct the distance-redshift relation based on the Pantheon dataset, extending it up to the maximum redshift of the available SGL systems. Deep learning is a realm dedicating to the research of various Artificial Neural Networks (ANN), which is composed of layers of neurons modeled after the biological neurons in human brain. Hence, deep learning is fantastic to deal with large and highly complex tasks, such as classification, clustering, generation and so on. Deep learning has emerged as a powerful tool in various cosmological research areas, demonstrating its effectiveness in tasks such as predicting galaxy morphology \citep{Dieleman:2015}, constraining dark energy \citep{Escamilla-Rivera:2020}, and calibrating Gamma-ray bursts (GRBs) \citep{Luongo:2020hyk, Tang:2022}. In our recent work \citep{Tang:2021}, we applied deep learning techniques to reconstruct the distance-redshift relation of SNe Ia without making any assumptions about the cosmological model or the parametric form of the relation. Furthermore, we utilized this reconstructed relation to investigate potential redshift dependencies in the luminosity corrections of GRBs. Unlike the GP method, which is constrained to reconstruct the curve within the data region, deep learning has the capacity to extend the reconstruction far beyond the available data region. Thus all of the SGL systems can be used and the constraint on the spatial curvature would be tighter.

The structure of the remaining sections of this paper is as follows: Section \ref{sec:Methodology} provides an overview of the DSR method and the lens mass models utilized in constraining the spatial curvature. Section \ref{sec:data} outlines the observational datasets employed in the analysis and details of the procedure for reconstructing the distance-redshift relation using deep learning techniques. The obtained results are presented in Section \ref{sec:Result}. Lastly, Section \ref{sec:summary} contains the discussion and summary.

\section{Methodology}\label{sec:Methodology}
In the context of a homogeneous and isotropic Universe, the spacetime can be described by the Friedmann-Robertson-Walker (FRW) metric, given by
\begin{equation}\label{eq:FRW}
ds^2=-c^2dt^2+\frac{a(t)^2}{1-Kr^2}dr^2+a(t)^2r^2d\Omega^2,
\end{equation}
where $c$ represents the speed of light, $K$ is a constant that denotes the spatial curvature of the Universe. Specifically, when $K<0$, $K=0$, and $K>0$, it corresponds to an open, flat, and closed Universe, respectively.
The scale factor $a(t)$ represents the expansion of the Universe with respect to cosmic time, and its derivative $\dot{a} \equiv \frac{da}{dt}$ defines the Hubble parameter $H \equiv \frac{\dot{a}}{a}$. To quantify the spatial separation between a source at redshift $z_s$ observed from redshift $z_l$, the dimensionless comoving distance is expressed as:
\begin{equation}
d(z_l,z_s)=\frac{1}{\sqrt{|\Omega_k|}}S_k\left(\sqrt{|\Omega_k|}\int^{z_s}{z_l}\frac{dz^{\prime}}{E(z^{\prime})}\right),
\end{equation}
where $\Omega_k \equiv -\frac{Kc^2}{H_0^2a_0^2}$ represents the normalized curvature parameter. The reduced Hubble parameter is denoted as $E(z) \equiv \frac{H(z)}{H_0}$, where $H_0$ represents the present-day value of the Hubble parameter. The function $S_k$ is defined as follows:
\begin{eqnarray}
S_{k}(x)=
\begin{cases}
\sinh(x),& (\Omega_k>0),\\
x,& (\Omega_k=0),\\
\sin(x),& (\Omega_k<0).
\end{cases}
\end{eqnarray}

For simplicity, we introduce the notation $d(z) \equiv d(0,z)$, $d_l \equiv d(0,z_l)$, $d_s \equiv d(0,z_s)$, and $d_{ls} \equiv d(z_l,z_s)$. Under the assumption that cosmic time $t$ and redshift $z$ have a one-to-one correspondence, and with the condition that the derivative of $d(z)$ with respect to $z$ satisfies $d'(z)>0$, the three dimensionless distances ($d_l$, $d_s$, and $d_{ls}$) are connected through the distance sum rule (DSR) relation \citep{Rasanen:2015}:

\begin{equation}\label{eq:DSR}
\frac{d_{ls}}{d_s}=\sqrt{1+\Omega_kd_l^2}-\frac{d_l}{d_s}\sqrt{1+\Omega_kd_s^2}.
\end{equation}
If the Universe is accurately described by the FRW metric, the curvature parameter $\Omega_k$ should be a constant. Therefore, if the validity of the FRW metric is confirmed, the DSR relation provides a means to constrain the value of $\Omega_k$. By analyzing the relation between the dimensionless distances, we can obtain valuable insights into the spatial curvature of the Universe.

The dimensionless comoving distances $d_l$ and $d_s$ can be obtained through the analysis of SNe Ia data. On the other hand, the distance ratio $d_{ls}/d_s$ is determined using the data from SGL observations.
In terms of the Einstein radius and the velocity dispersion associated with the lens mass profile, the expression for the distance ratio can be formulated. For certain gravitational lens systems, the mass distribution of the lens has been observed to closely approximate an isothermal profile \citep{Cohn:2000ds, Munoz:2001bw, Rusin:2001ce, Treu:2002ee, Rusin:2004kq}. Consequently, the Singular Isothermal Sphere (SIS) model has emerged as a prevalent and straightforward choice for describing the lens mass profile. This model effectively emulates the flat rotation curves characteristic of galaxies, featuring a density inversely proportional to the square of the galaxy's radius. Additionally, the structure of galaxies has been extensively explored through N-body simulations \citep{Navarro:1995iw, Moore:1997sg}. Navarro et al. \citep{Navarro:1995iw} discovered that the halo profile of a galaxy exhibits an approximate isothermal behavior across a wide range of radii. However, it deviates from the $r^{-2}$ power-law near the central region, transitioning to a steeper profile than $r^{-2}$ as the distance approaches the virial radius. To adequately address the specific features of the halo profile, a density function incorporating a power-law dependence on radius has been introduced to describe the lens profile. This density function resembles the generalized Navarro-Frenk-White (NFW) profile \citep{Navarro:1995iw}. Among these descriptions, the Power-Law (PL) model stands out, characterized by a variable power-law index denoted as $\gamma$. Remarkably, when $\gamma$ is set to 2, the profile aligns with a configuration akin to a singular isothermal model. It is noteworthy that these profile descriptions do not inherently distinguish between luminosity density and total mass density. When accounting for the presence of dark matter, the luminosity density may diverge from the overall galaxy profile. This prompts the introduction of models like the Extended Power-Law (EPL) model, which accommodates the complexities stemming from both luminosity density and dark matter distribution within the lens mass. Hence, we consider three distinct lens models: the Singular Isothermal Sphere (SIS) model, the Power-Law (PL) model, and the Extended Power-Law (EPL) model, to comprehensively investigate the influence of various lens galaxy mass profiles on the constraints imposed on the spatial curvature.

Within the singular isothermal sphere (SIS) model, the mass density distribution of the lens galaxy follows a scaling relation of $\rho\propto r^{-2}$. This leads to an expression for the distance ratio as follows \citep{Mollerach:2002}:
\begin{equation}\label{eq:SIE}
\frac{d_{ls}}{d_s}=\frac{c^2\theta_{\rm{E}}}{4\pi\sigma_{\rm{SIS}}^2},
\end{equation}
where $\theta_E$ represents the Einstein radius, and $\sigma_{\rm{SIS}}$ is the velocity dispersion associated with the lens mass profile. It is worth noting that the equivalence between the observed stellar velocity dispersion $\sigma_0$ and $\sigma_{\rm{SIS}}$ within the context of the SIS model is not an absolute requirement \citep{Khedekar2011}. There may be potential deviations between the observed stellar velocity dispersion and the characteristic velocity dispersion associated with the SIS model. Consequently, to account for such deviation, a phenomenological parameter $f$ is introduced, yielding $\sigma_{\rm{SIS}}=f\sigma_0$ \citep{Kochanek:1992,Ofek:2003,Cao:2012,Li:2019}. Notably, the free parameter $f$ is anticipated to fall within the range of $0.8<f^2<1.2$ \citep{Ofek:2003}. In practice, the velocity dispersion is typically measured within the aperture radius $\theta_{ap}$ in actual SGL data. To convert the measurement to $\sigma_0$, an aperture correction formula \citep{Jorgensen:1995} can be employed, given by the equation
\begin{equation}
\sigma_0=\sigma_{\rm ap}\left(\frac{\theta_{\rm{eff}}}{2\theta_{\rm{ap}}}\right)^{\eta},
\end{equation}
where $\sigma_{\rm ap}$ represents the luminosity weighted average of the line-of-sight velocity dispersion within the aperture radius, $\theta_{\rm{eff}}$ corresponds to the effective angular radius, and $\eta$ is the correction factor fixed to $-0.066$ \citep{Cappellari:2006,Chen:2019}. It is important to consider that the uncertainty associated with $\sigma_{\rm ap}$ propagates to $\sigma_0$, subsequently impacting $\sigma_{\rm{SIS}}$. Additionally, the uncertainty in the distance ratio $d_{ls}/d_s$ is derived from the uncertainties in $\theta_E$ and $\sigma_{\rm{SIS}}$. In this work, we adopt a fractional uncertainty of $5\%$ for $\theta_E$ \citep{Liao:2015uzb}.

Within the framework of the power-law spherical (PL) model, the mass density distribution of the lensing galaxy is characterized by a spherically symmetric power-law behavior, expressed as $\rho\propto r^{-\gamma}$, where $\gamma$ represents the power-law index. The distance ratio in PL model can be described as \citep{Koopmans:2006}
\begin{equation}\label{eq:PLS}
\frac{d_{ls}}{d_s}=\frac{c^2\theta_{\rm{E}}}{4\pi\sigma^2_{\rm{ap}}}\left(\frac{\theta_{\rm ap}}{\theta_{\rm E}}\right)^{2-\gamma}f^{-1}(\gamma),
\end{equation}
where
\begin{equation}
f(\gamma)=-\frac{1}{\sqrt{\pi}}\frac{(5-2\gamma)(1-\gamma)}{3-\gamma}\frac{\Gamma(\gamma-1)}{\Gamma(\gamma-3/2)}\left[\frac{\Gamma(\gamma/2-1/2)}{\Gamma(\gamma/2)}\right]^2.
\end{equation}
It is worth noting that when $\gamma$ takes the value of 2, the PL model reduces to the standard SIS model. To account for the potential redshift evolution of the mass density profile, we introduce a parameterization for $\gamma$, expressed as $\gamma(z_l)=\gamma_0+\gamma_1z_l$, where $\gamma_0$ and $\gamma_1$ represent two independent free parameters.

Within the extended power-law (EPL) model, the luminosity density profile $\nu(r)$ can differ from the total mass density profile $\rho(r)$, accounting for the presence of a dark matter halo. We adopt the following functional forms for the power-law mass density profile and the luminosity density of stars, respectively:
\begin{equation}
\rho(r)=\rho_0\left(\frac{r}{r_0}\right)^{-\alpha}, \quad \nu(r)=\nu_0\left(\frac{r}{r_0}\right)^{-\delta},
\end{equation}
where $\alpha$ and $\delta$ correspond to the power-law index parameters, $r_0$ represents the characteristic length scale, and $\rho_0$ and $\nu_0$ are normalization constants. The distance ratio in the EPL model is expressed as \citep{Birrer:2018vtm,Lee:2021jcg}:
\begin{equation}\label{eq:EPL}
\frac{d_{ls}}{d_s}=\frac{c^2\theta_{\rm{E}}}{2\sigma_0^2\sqrt{\pi}}\frac{3-\delta}{(\xi-2\beta)(3-\xi)}\left(\frac{\theta_{\rm eff}}{\theta_E}\right)^{2-\alpha}\left[\frac{\lambda(\xi)-\beta\lambda(\xi+2)}{\lambda(\alpha)\lambda(\delta)}\right],
\end{equation}
where $\xi=\alpha+\delta-2$, $\lambda(x)=\Gamma\left(\frac{x-1}{2}\right)/\Gamma\left(\frac{x}{2}\right)$, and $\beta$ represents an anisotropy parameter that characterizes the anisotropic distribution of the three-dimensional velocity dispersion. In accordance with \cite{Wang:2020}, we consider $\beta$ as a nuisance parameter and marginalize over it with a Gaussian prior of $\beta=0.18\pm0.13$. Simultaneously, we treat $\alpha$ and $\delta$ as free parameters. It is worth noting that when $\alpha=\delta=2$ and $\beta=0$, the EPL model reduces to the standard SIS model.

\section{Observational data and deep learning}\label{sec:data}

\begin{figure}[htbp]
    \centering
    \includegraphics[width=0.95\textwidth]{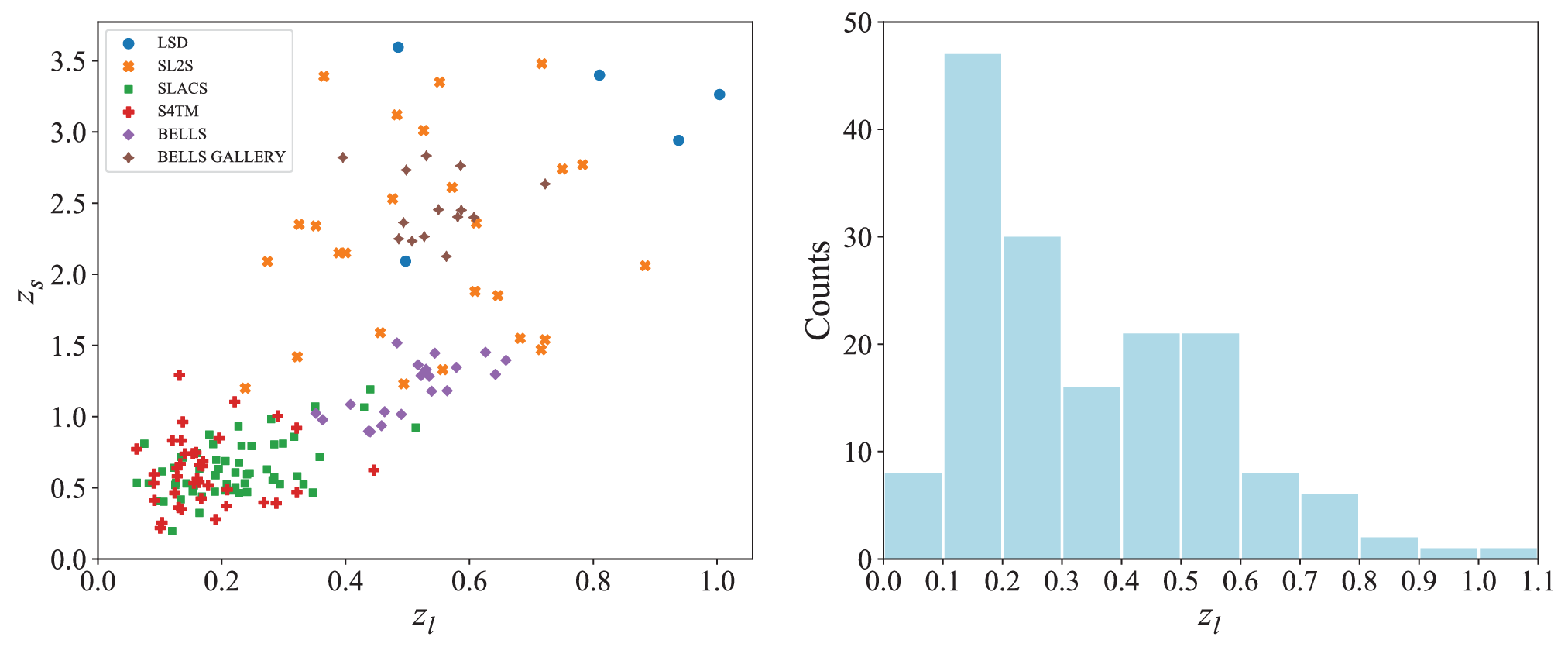}
    \caption{\small{Left: the distribution of the SGL sample obtained from various surveys in the $z_l$-$z_s$ plane. Right: the redshift distribution of lens.}}
    \label{fig:distrib}
\end{figure}

The distance ratios $d_{ls}/d_s$ are obtained from the observations of strong gravitational lensing (SGL) systems. In a recent study, \cite{Chen:2019} compiled a new SGL sample by combining data from various galaxy surveys, including the Lenses Structure and Dynamics (LSD) survey \citep{TreuandKoopmans:2004}, the Sloan Lens ACS (SLACS) survey \citep{Bolton:2006}, the CFHT Strong Lensing Legacy Survey \citep{Cabanac:2007}, and the BOSS Emission-Line Lens Survey \citep{Brownstein:2011}. This compiled sample consists of 161 galaxy-scale SGL systems, covering a redshift range of $z_l\in[0.0624,1.004]$ for the lens galaxies and $z_s\in[0.197,3.595]$ for the source galaxies. In Figure \ref{fig:distrib}, we illustrate the distribution of the SGL sample, derived from diverse survey sources, as depicted in the $z_l$-$z_s$ plane. Additionally, we provide the redshift distribution of the lens objects, with a predominant concentration of lenses residing at an approximate redshift of $z_l\sim0.2$.

The dimensionless comoving distances $d_l$ and $d_s$ are derived from the luminosity distance $D_L$ of SNe Ia using the relation
\begin{equation}\label{eq:d_DL}
d(z)=\frac{H_0D_L(z)}{c(1+z)}.
\end{equation}
The luminosity distance $D_L$ can be obtained from the light curve of SNe Ia. Considering a specific redshift $z$, the distance modulus of SNe Ia can be expressed as
\begin{equation}\label{eq:muDL}
\mu=5\log_{10}\frac{D_L(z)}{\textrm{Mpc}}+25=m_{B, \textrm{corr}}-M_B,
\end{equation}
where $M_B$ represents the absolute magnitude and $m_{B, \textrm{corr}}$ denotes the corrected apparent magnitude observed in the B-band, reported in the largest and most recent Pantheon dataset \citep{Scolnic:2018}. The redshift range of SNe Ia sample used in our work , i.e. the Pantheon dataset, is $z\in[0.01,2.30]$.

To obtain the comoving distances $d$ at the redshifts of the lens and source for all the SGL systems, it is necessary to reconstruct a continuous curve of the distance-redshift relation $d(z)$ based on the Pantheon sample. Previous work by \cite{Wang:2020} employed the Gaussian Process (GP) method to reconstruct a smooth curve of $d(z)$ from SNe Ia data. However, the reconstructed uncertainty of the GP method tends to be large in regions where the data points are sparse, and it becomes even more challenging to estimate distances beyond the observed redshift range. Consequently, SGL systems with source redshifts larger than 2.3 could not be utilized in their analysis.

In this paper, we adopt a deep learning method to reconstruct the distance-redshift curve without any specific assumption about its parametric form. This approach allows us to reconstruct the distance-redshift relation using a wide range of redshifts, covering the entire redshift range of the SGL sample. Specifically, we can extend the reconstruction up to a redshift of $z=4$, thus ensuring that we encompass the full redshift range of the SGL systems under consideration. This utilization of deep learning enables us to overcome the limitations associated with the sparse data points and extrapolate the distance-redshift relation to regions beyond the direct observational range.

Deep learning has emerged as a powerful methodology for analyzing complex and intricate datasets. One common approach involves the utilization of Artificial Neural Networks (ANNs) as underlying models, such as Convolutional Neural Networks (CNNs), Recurrent Neural Networks (RNNs), Bayesian Neural Networks (BNNs), among others. These neural networks typically consist of multiple layers of interconnected processing units, where each layer receives information from the previous layer and transforms it to the subsequent layer. Through training, these networks aim to learn and represent the underlying patterns and structures within the data.
In the context of our research, we employ RNNs as a key component of our deep learning approach. RNNs are well-suited for handling sequential data and making predictions based on learned data representations. By feeding the Pantheon dataset into the RNN, we can effectively capture the relationship between the distance modulus $\mu$ and the redshift $z$. This enables us to predict distances at arbitrary redshifts, even beyond the range covered by the observational data. However, RNNs alone are insufficient for providing uncertainty estimates for these predictions. To address this limitation, we incorporate BNNs into our network architecture. BNNs serve as a complementary component to the RNNs and allow us to calculate the uncertainty associated with the distance predictions. Our previous work \citep{Tang:2021} had incorporated both RNNs and BNNs to model the distance modulus-redshift relationship based on the Pantheon dataset, while this current research emphasizes the reconstruction of the distance curve $d(z)$ using the deep learning approach.

\begin{figure}[htbp]
\centering
\includegraphics[width=0.8\textwidth]{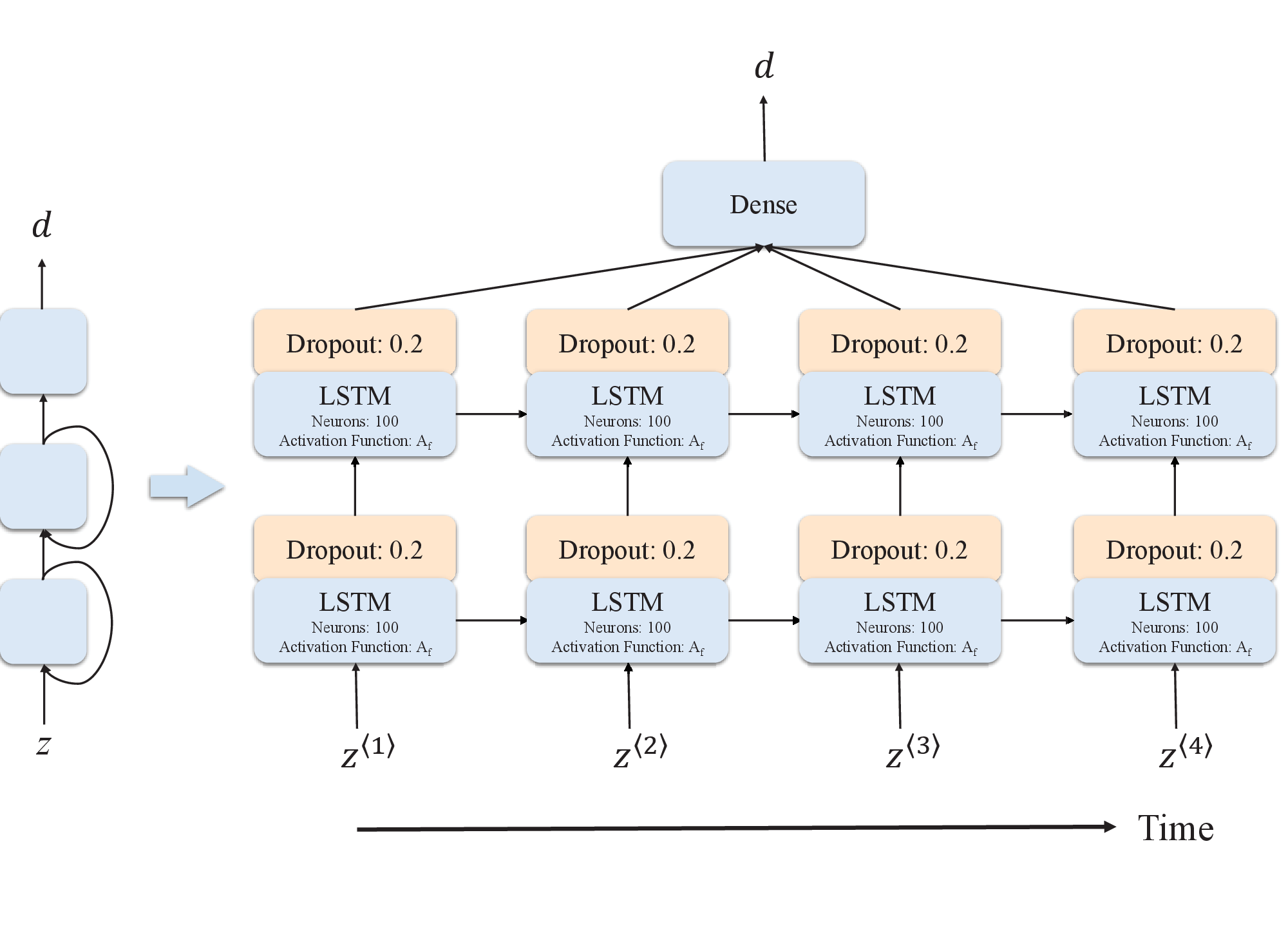}
\caption{\small{Left: The network architecture comprising a single hidden layer is illustrated. Right: The network unfolded up to time step $t=4$, denoted as $Z^{\langle i\rangle}$ representing the $i$th time step. In our network, both the input layer and hidden layer are composed of LSTM cells, housing 100 neurons each. The output layer is a fully-connected (dense) layer. To mitigate overfitting, the dropout technique is implemented between each LSTM cell and its subsequent layer.}}\label{fig:d_RNN}
\end{figure}

The architecture of our network is illustrated in Figure \ref{fig:d_RNN}. The central component is the RNN, which consists of three layers: an input layer that receives the redshift $z$ as the feature, a hidden layer that processes information from the previous layer and passes it to the next layer, and an output layer that generates the target output, which in this case is the comoving distance $d$. The RNN is designed to capture the temporal dependencies and patterns in the input data.To overcome the challenges associated with training RNNs on long sequential data and to address the issue of information retention over long periods, we employ Long Short-Term Memory (LSTM) cells as the basic units of our network. LSTM cells enhance RNNs by incorporating explicit memory mechanisms, allowing the network to selectively store, discard, and retrieve information. The input and hidden layers of our network consist of 100 LSTM cells each.

In the training process, the RNN is fed with the Pantheon data to learn and represent the relationship between the comoving distance $d$ and the redshift $z$. This is achieved by minimizing a loss function that quantifies the discrepancy between the network's predictions and the observed distances. In this work, we utilize the mean-squared-error (MSE) function as the loss function, and we employ the Adam optimizer to find the minimum of this function. To enhance the network's performance, we introduce a non-linear activation function denoted as $A_f$. In our previous research on reconstructing the distance modulus $\mu(z)$, we found that the hyperbolic tangent (tanh) function outperformed other activation functions such as ReLU, ELU, and SELU. However, since we are now reconstructing the comoving distance $d(z)$ instead of the distance modulus, we compare the performance of all four activation functions (tanh, ReLU, ELU, and SELU) to determine the most suitable choice. By setting the time step to $t=4$ and utilizing the LSTM-based RNN architecture with appropriate activation functions, our network aims to learn the complex relationship between the redshift $z$ and the comoving distance $d$. Through training and optimization, we obtain a model that can predict distances at arbitrary redshifts, including those beyond the range covered by the Pantheon dataset.

In the context of BNN, it is worth noting that designing a traditional BNN is a challenging task due to its inherent complexity. Fortunately, \cite{Gal:2016a,Gal:2016b,Gal:2016c} had demonstrated that dropout, commonly used in deep neural networks as a regularization technique to address the issue of overfitting, can be viewed as an approximation to Bayesian inference in deep Gaussian processes. This means that a network incorporating dropout can be considered mathematically equivalent to a Bayesian model. In this study,  we incorporate the dropout technique within the RNN to emulate the characteristics of a BNN. By executing the trained network multiple times, we can generate multiple predictions for the comoving distance at different redshifts. This process allows us to obtain a range of possible predictions and, consequently, estimate the confidence region associated with these predictions. This approach effectively mimics the behavior of a BNN, where the network models the posterior distribution over the parameters. In our research, we employ a dropout rate of 0.2 between the LSTM layer and its subsequent layer.

To begin the reconstruction of the comoving distance $d(z)$, we first normalize the comoving distance data obtained from the Pantheon compilation according to equations (\ref{eq:d_DL}) and (\ref{eq:muDL}) with the chosen parameters $H_0 = 70$ km s$^{-1}$ Mpc$^{-1}$ and $M_B = -19.36$ \citep{Scolnic:2014}. Next, we sort the normalized data points $(z_i, d_{i})$ in ascending order of redshift $z_i$ and reorganize them into four sequences. In each sequence, the redshifts and the corresponding normalized distances are used as input and output vectors, respectively, for training the network. Subsequently, we train the network constructed as described above using TensorFlow\footnote{https://www.tensorflow.org} for a total of 1000 iterations. The well-trained network is saved for later use. In the final step, we execute the trained network 1000 times to predict the distance $d$ over the redshift range $z \in [0, 4]$. The distribution of the predicted distances is obtained as a Gaussian distribution.

The results of the distance reconstruction using the four activation functions (tanh, ReLU, ELU, and SELU) are plotted in Figure \ref{fig:sim}. For comparison, we also include the best-fitting curve of the $\Lambda$CDM model (represented by the black line). It is worth noting that while the uncertainty in the reconstructed curve using deep learning may be slightly larger than that obtained using the GP method within the data region, the advantage of deep learning lies in its ability to reconstruct the curve beyond the data region. This enables us to leverage the full sample of SGL systems.

\begin{figure}[htbp]
\label{fig:sim}
\centering
\includegraphics[width=0.48\textwidth]{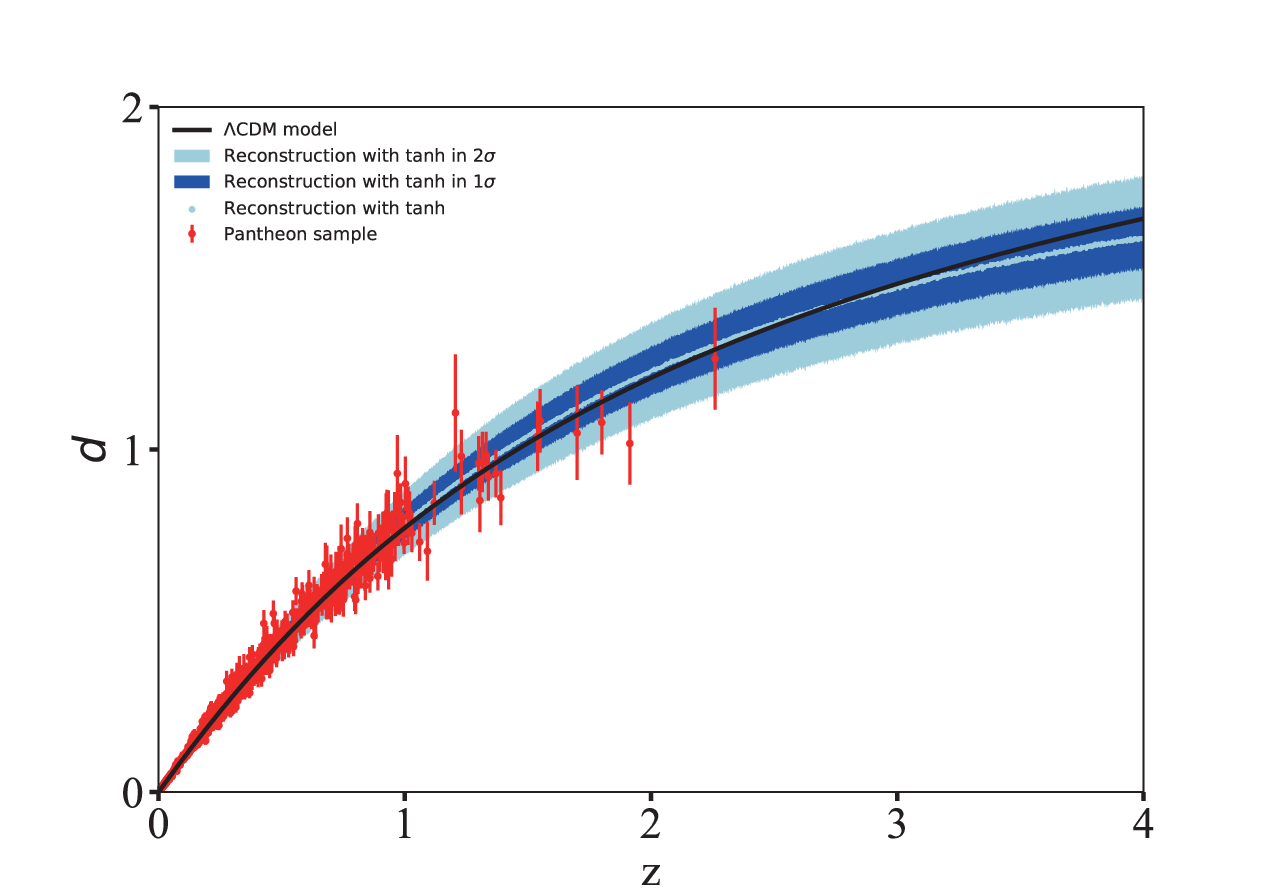}
\includegraphics[width=0.48\textwidth]{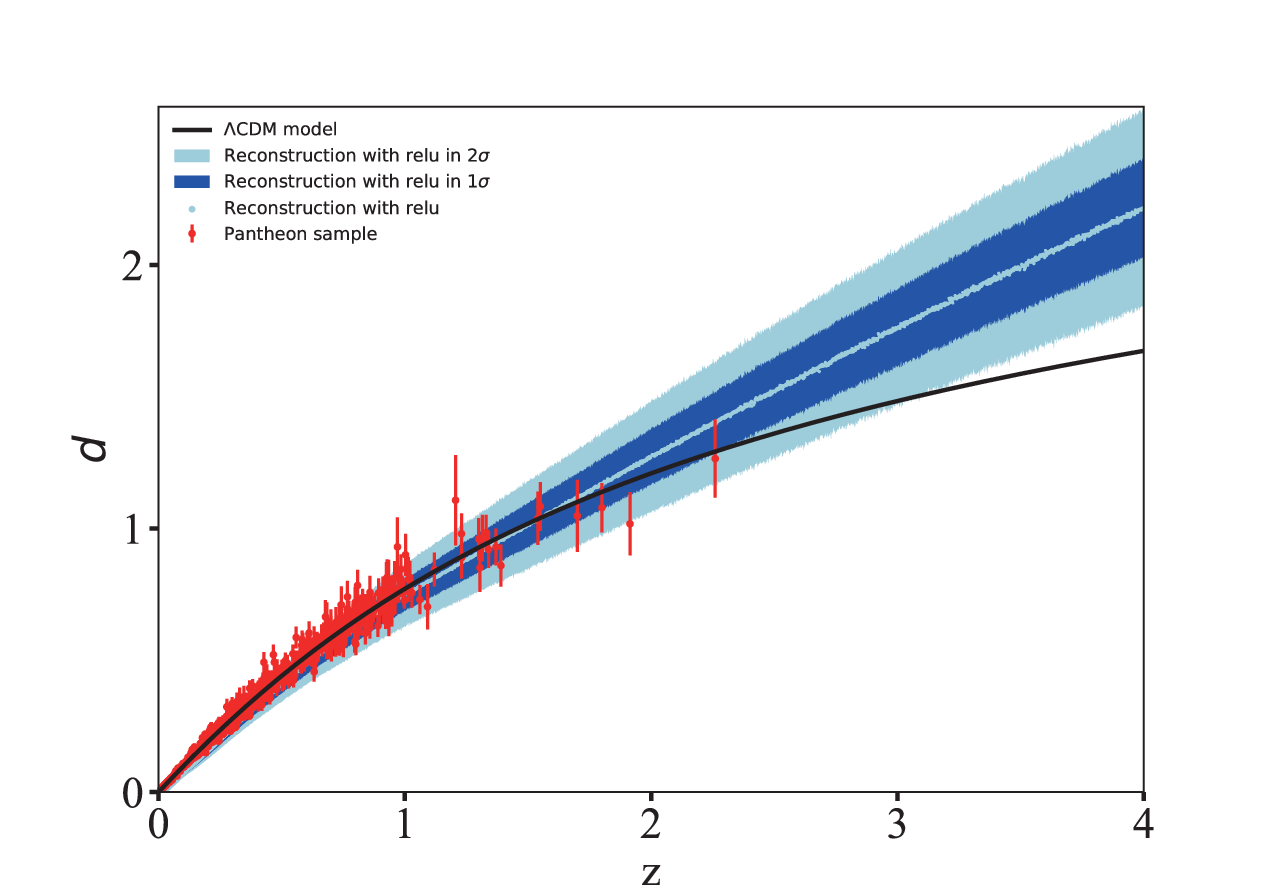}
\includegraphics[width=0.48\textwidth]{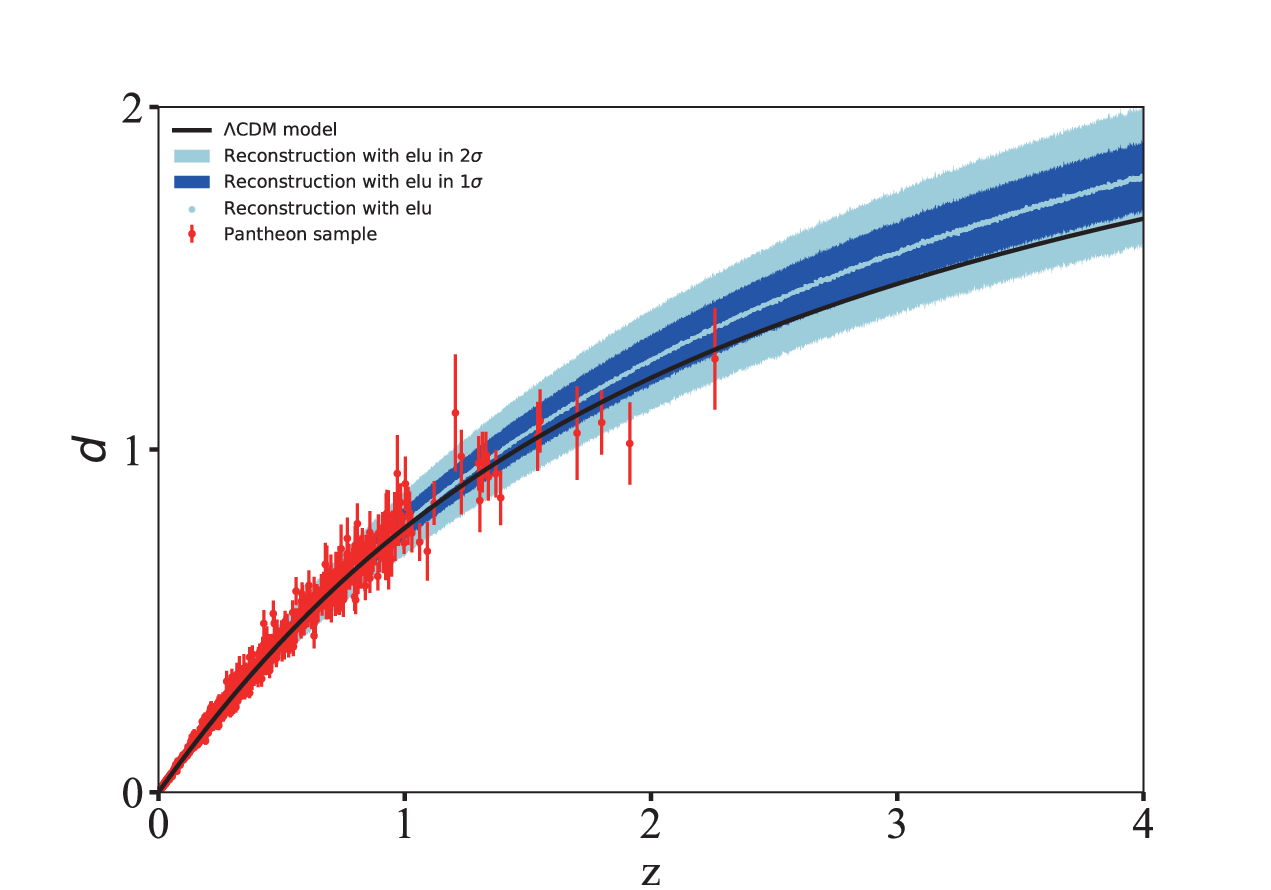}
\includegraphics[width=0.48\textwidth]{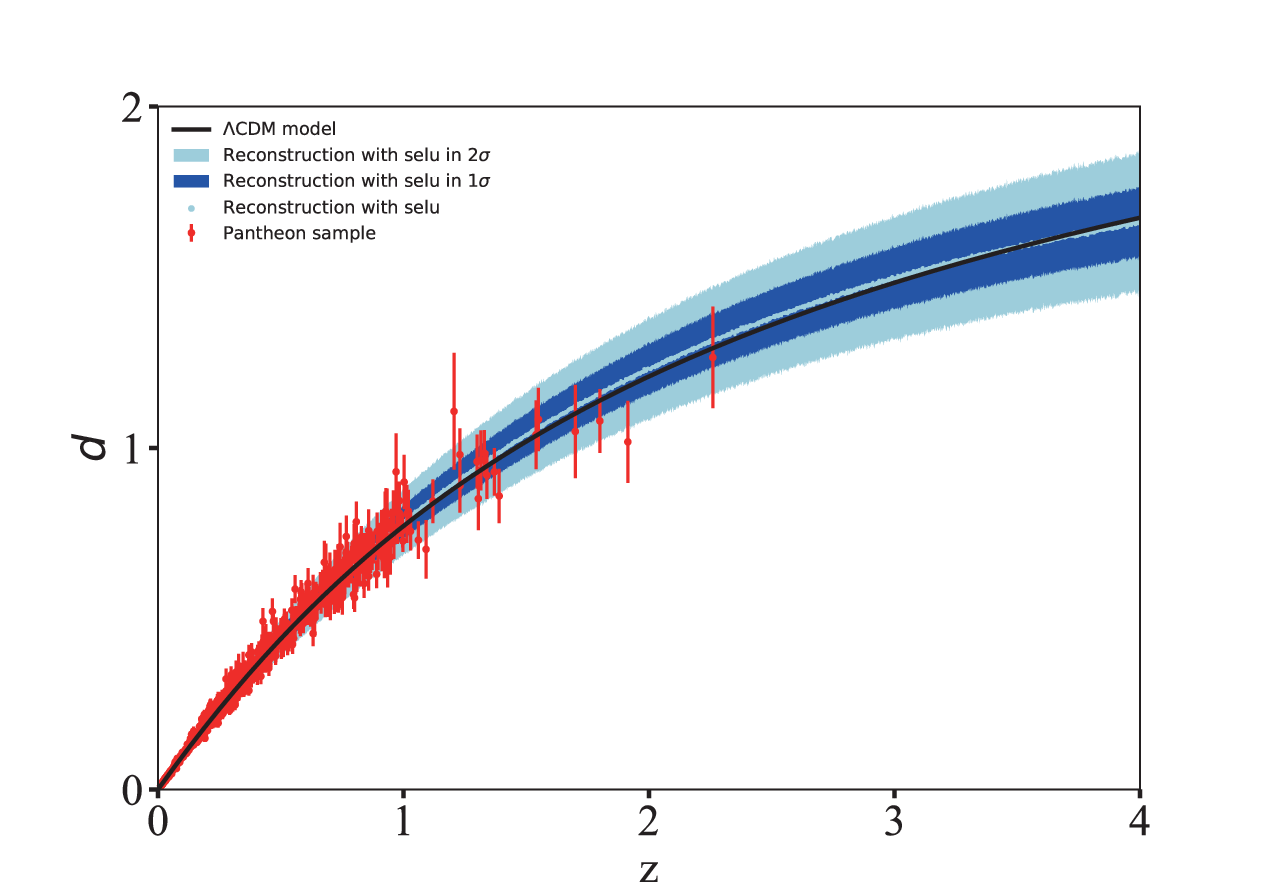}
\caption{\small{The reconstructions of $d(z)$ from the Pantheon data are presented, employing four distinct activation functions. Top-left: tanh; top-right: relu; bottom-left: elu; bottom-right: selu.}}\label{fig:sim}
\end{figure}

As depicted in the results, it is observed that only the reconstructions using the tanh and selu activation functions are consistent with the flat $\Lambda$CDM model within the 1$\sigma$ confidence level. Considering that most of the current cosmological probes favor the LCDM model, and the reconstructed curves using relu and elu functions deviate from the $\Lambda$CDM model too much at high redshift, these two activation functions are excluded in the following calculation. Therefore, we will derive the dimensionless comoving distances of the SGL systems from the reconstruction with the tanh and selu functions, respectively. We emphasize that the reconstructed curves using deep learning are independently of cosmological model. The $\Lambda$CDM curves plotted in Figure \ref{fig:sim} are just for comparison.

\section{Results}\label{sec:Result}

With the reconstructed $d(z)$ curve, we can obtain the dimensionless comoving distance and the corresponding uncertainty at $z_l$ and $z_s$, then calculate the distance ratio $R_{\rm SNe}\equiv d_{ls}/d_s$ according to equation (\ref{eq:DSR}), and the uncertainty $\sigma_{R_{\rm SNe}}$ propagates from the uncertainties of $d_l$ and $d_s$. Besides, the distance ratios $R_{\rm SGL}\equiv d_{ls}/d_s$ can also be obtained from SGL system using equations (\ref{eq:SIE}), (\ref{eq:PLS}) or (\ref{eq:EPL}), according to different mass models of lens galaxy. The corresponding uncertainty $\sigma_{R_{\rm SGL}}$ propagates from the uncertainties of the observations of SGL.
To compare the distance ratio obtained from SNe Ia and SGL systems, we determine the best-fitting parameters by maximizing the likelihood function, which is proportional to the exponential of the negative chi-square statistic, i.e. $\mathcal{L}\propto \exp({-\chi^2/2})$, where
\begin{equation}
\chi^2(\bm{p}, \Omega_k) = \sum_{i=1}^{161} \frac{\left(R_{\rm SNe} - R_{\rm SGL}\right)^2}{\sigma_{\rm total}^2},
\end{equation}
Here, $\bm{p}$ represents the set of parameters for the lens mass profile, where $\bm{p}=f$ for the SIS model, $\bm{p}=(\gamma_0, \gamma_1)$ for the PL model, and $\bm{p}=(\alpha, \delta)$ for the EPL model. The term $\sigma_{\rm total}$ represents the total uncertainty, which includes contributions from the uncertainty in the reconstruction and the uncertainty propagated from the SGL observations:

\begin{equation}
\sigma_{\rm total}^2 = \sigma_{R_{\rm SNe}}^2 + \sigma_{R_{\rm SGL}}^2.
\end{equation}

Assuming a flat prior on all free parameters, we calculate the posterior Probability Density Function (PDF) of the parameter space using the Python package \textsf{emcee} \citep{Foreman-Mackey:2013}. It is worth noting that the prior on the spatial curvature parameter $\Omega_k$ is set to $\Omega_k \geq -0.39$ to ensure that $1+\Omega_kd_l^2 \geq 0$ and $1+\Omega_kd_s^2 \geq 0$ within the redshift range $z \leq 4$.

\begin{table}[htbp]
\centering
\caption{\small{The best-fitting parameters in the framework of SIS model using the distance reconstructed with tanh and selu functions.}}\label{tab:paras_SIS}
\arrayrulewidth=1.0pt
\renewcommand{\arraystretch}{1.3}
\begin{tabular}{c|cc} 
\hline\hline 
 & $\Omega_k$     & $f$   \\
\hline
tanh  &$0.049^{+0.147}_{-0.125}$	&$1.038^{+0.009}_{-0.008}$\\
selu  &$0.082^{+0.125}_{-0.130}$	&$1.039^{+0.009}_{-0.009}$\\
\hline
\end{tabular}
\end{table}

\begin{figure}[htbp]
    \centering
    \includegraphics[width=0.6\textwidth]{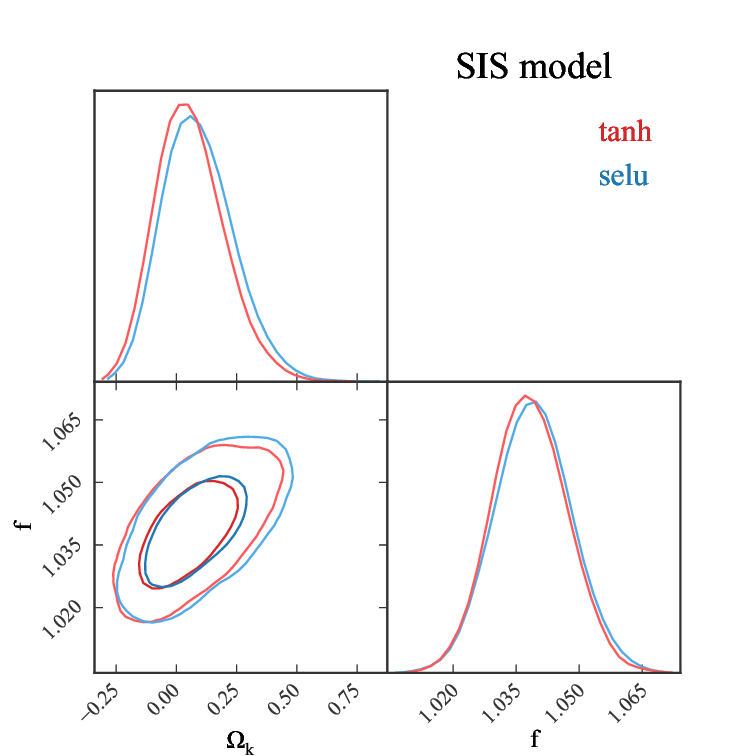}
    \caption{\small{The 2-dimensional confidence contours and 1-dimensional PDFs for the parameters within the SIS model framework are depicted. The results obtained using the distance reconstructed with the tanh and selu are represented by the red and blue lines, respectively.}}
    \label{fig:SIS}
\end{figure}

In the context of the SIS lens model, we present the best-fitting parameters obtained using the reconstructions with the tanh and selu activation functions in Table \ref{tab:paras_SIS}. Additionally, we provide the 1$\sigma$ and 2$\sigma$ confidence contours as well as the marginalized PDFs for the parameter space in Figure \ref{fig:SIS}. For the spatial curvature, it is constrained to be $\Omega_k=0.049^{+0.147}_{-0.125}$ with tanh function and $\Omega_k=0.082^{+0.125}_{-0.130}$ with selu function. The constraints on $\Omega_k$ both in two function support a flat Universe within 1$\sigma$ confidence level, consistent with the Planck results \citep{PlanckCollaboration:2020}. The constraint on the parameter $f$ is rather tight, $1.038^{+0.009}_{-0.008}$ with tanh function and $1.039^{+0.009}_{-0.009}$ with selu function. Both of them exclude the standard SIS model ($f=1$) at more than 4$\sigma$ confidence level. This indicates that the lens mass profile slightly but with strong evidence deviates from the standard SIS model.

\begin{table}[htbp]
\centering
\caption{\small{The best-fitting parameters in the framework of PL model using the distance reconstructed with tanh and selu functions.}}\label{tab:paras_PL}
\arrayrulewidth=1.0pt
\renewcommand{\arraystretch}{1.3}
{\begin{tabular}{c|ccc} 
\hline\hline 
 & $\Omega_k$     & $\gamma_0$   &$\gamma_1$\\
\hline
tanh  &$-0.245^{+0.075}_{-0.071}$	&$2.076^{+0.028}_{-0.030}$	 &$-0.309^{+0.114}_{-0.091}$\\
selu  &$-0.232^{+0.076}_{-0.068}$	&$2.074^{+0.028}_{-0.029}$	 &$-0.307^{+0.123}_{-0.093}$\\
\hline
\end{tabular}}
\end{table}

\begin{figure}[htbp]
    \centering
    \includegraphics[width=0.6\textwidth]{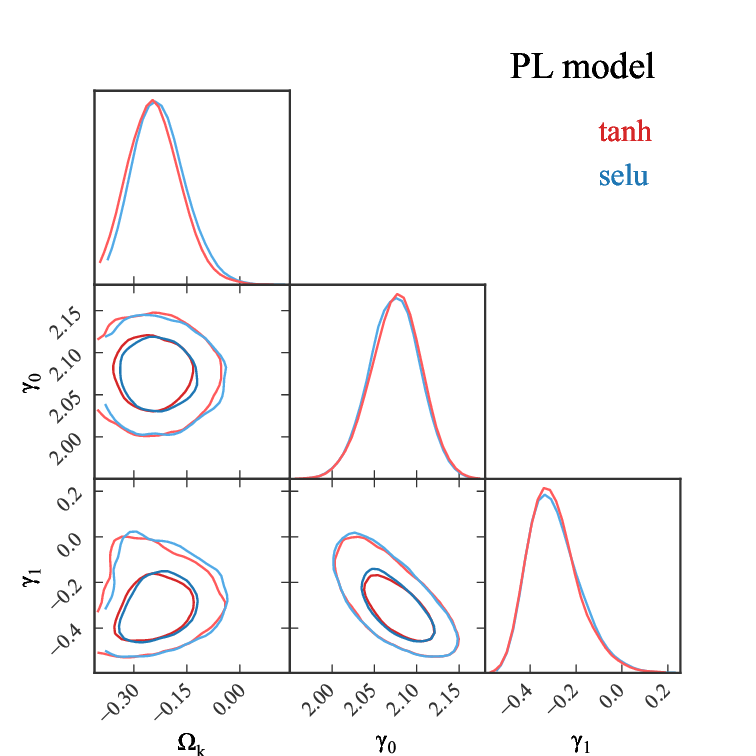}
    \caption{\small{The 2-dimensional confidence contours and 1-dimensional PDFs for the parameters within the PL model framework are depicted. The results obtained using the distance reconstructed with the tanh and selu are represented by the red and blue lines, respectively.}}
    \label{fig:model PL}
\end{figure}

In the context of the PL lens model, the parameters are presented in Table \ref{tab:paras_PL}. Additionally, the contours and PDFs for the parameter space are plotted in Figure \ref{fig:model PL}. Similar to the SIS model, the constraints with tanh and selu activation functions are consistent with each other at 1$\sigma$ confidence level. However, the constraint on curvature parameter in PL model is totally different from that in SIS model. The spatial curvature is constrained to be $\Omega_k=-0.245^{+0.075}_{-0.071}$ with tanh function and $\Omega_k=-0.232^{+0.076}_{-0.068}$ with selu function. The constraints on $\Omega_k$ in PL model prefer a closed Universe at $\sim 3\sigma$ confidence level. For the lens parameters, they are constrained to be $(\gamma_0,\gamma_1)=(2.076^{+0.028}_{-0.030},-0.309^{+0.114}_{-0.091})$ with tanh function, and $(\gamma_0,\gamma_1)=(2.074^{+0.028}_{-0.029},-0.307^{+0.123}_{-0.093})$ with selu function. The results deviate from the standard SIS model ($\gamma_0=2$, $\gamma_1=0$) at more than 2$\sigma$ confidence level, demonstrating that the total mass density profile of the lens galaxy possibly evolves with cosmic time.

\begin{table}[htbp]
\centering
\caption{\small{The best-fitting parameters in the framework of EPL model using the distance reconstructed with tanh and selu functions, the constraints of $\Omega_k$ are shown with the 95$\%$ confidence level upper limits.}}\label{tab:paras_EPL}
\arrayrulewidth=1.0pt
\renewcommand{\arraystretch}{1.3}
{\begin{tabular}{c|ccc} 
\hline\hline 
 & $\Omega_k$     & $\alpha$   &$\delta$\\
\hline
tanh  &$<0.011$	&2.114$^{+0.016}_{-0.019}$	 &2.383$^{+0.128}_{-0.099}$\\
selu  &$<0.051$	&2.112$^{+0.017}_{-0.018}$	 &2.375$^{+0.125}_{-0.089}$\\
\hline
\end{tabular}}
\end{table}

\begin{figure}[htbp]
\centering
\includegraphics[width=0.6\textwidth]{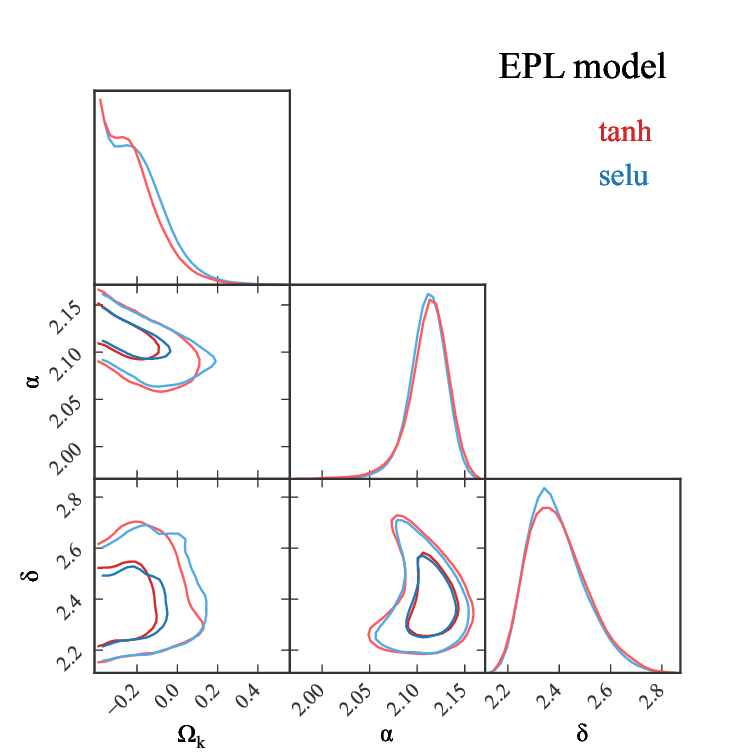}
\caption{\small{The 2-dimensional confidence contours and 1-dimensional PDFs for the parameters within the EPL model framework are depicted. The results obtained using the distance reconstructed with the tanh and selu are represented by the red and blue lines, respectively.}}\label{fig:EPL}
\end{figure}

In the context of the EPL lens model, the results obtained with two activation functions are shown in Table \ref{tab:paras_EPL} and Figure \ref{fig:EPL}. The constrain in EPL model is looser than that in SIS and PL models. With two functions, the spatial curvature parameters are constrained to be $\Omega_k<0.011$ in tanh function and $\Omega_k<0.051$ in selu function at $95\%$ confidence level, respectively. For the set of lens parameters, the results obtained with two activation functions are consistent with each other. We obtain $(\alpha,\delta)=(2.114^{+0.016}_{-0.019},2.383^{+0.128}_{-0.099})$ with tanh function and $(\alpha,\delta)=(2.112^{+0.017}_{-0.018},2.375^{+0.125}_{-0.089})$ with selu function. Both results deviate from the SIS model ($\alpha=\delta=2$) at more than 1$\sigma$ confidence level. Especially for $\alpha$, it rules out $\alpha=2$ at approximately 3$\sigma$ confidence level. These results show that the influence of the dark matter in the early-type galaxies should be considered and the total-mass profiles are not necessary to be consistent with the luminosity profiles.

\begin{figure}[htbp]
\centering
\includegraphics[width=0.7\textwidth]{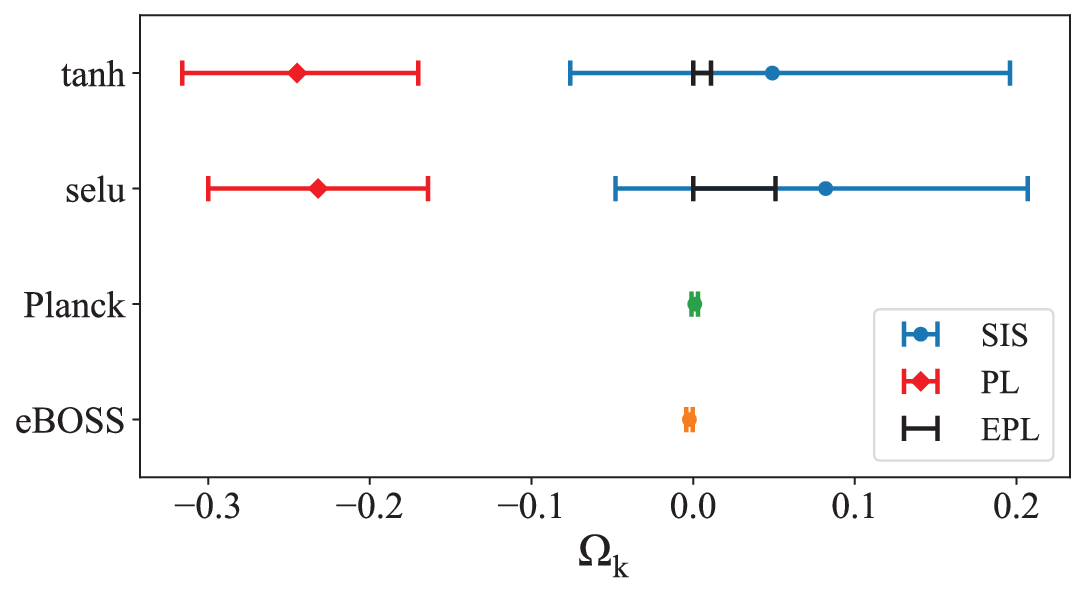}
\caption{\small{Constraint results of $\Omega_k$ in three different lens models using two activation functions in our work, compared with the constraints from other cosmological probes, Planck and eBOSS.}}\label{fig:compare}
\end{figure}

For enhanced clarity, Figure \ref{fig:compare} showcases the optimal fitting outcomes concerning the curvature parameter $\Omega_k$, accompanied by their corresponding 1$\sigma$ uncertainties within the context of the SIS and PL models. Additionally, the upper and lower bounds of $\Omega_k$ within the EPL model are exhibited. Furthermore, to facilitate comprehensive comparison, we integrate the constraints on $\Omega_k$ originating from alternative cosmological methodologies, including outcomes from the Planck \citep{PlanckCollaboration:2020} and the extended Baryon Oscillation Spectroscopic Survey (eBOSS) \citep{eBOSS:2020yzd}. Upon meticulous scrutiny, it becomes conspicuous that the constraints derived from the PL model exhibit noteworthy deviations from the outcomes of the Planck and eBOSS investigations. Meanwhile, the results stemming from the SIS and EPL models manifest congruence with the findings of the Planck and eBOSS initiatives. It is worth noting that the constraints associated with the EPL model, while consistent, display a marginally reduced stringency. Our findings underscore that, if the Universe is indeed flat, subtle deviations from the isothermal profile are discernible within the lens distribution. Moreover, it is imperative to duly consider variables such as the redshift evolution of lens profiles and the intricate interplay of dark matter in the broader landscape of cosmological research.

\section{Discussion and Summary}\label{sec:summary}

Based on geometrical optics, the distance sum rule (DSR) offers a model-independent approach to testing the validity of the FRW metric in cosmology. The DSR has proven to be a valuable tool for constraining the spatial curvature of the Universe.  Applying the DSR method, \cite{Wang:2020} recently investigated the spatial curvature with the combination of a SGL sample and the latest Pantheon SNe Ia. Although the total number of SGL systems is 161, the available SGL systems in \cite{Wang:2020} is just 135 due to that the GP regression used to reconstruct the distance-redshift relation can not reconstruct the curve well beyond the data region. In this research, we use the same data samples but with deep learning method to constrain the spatial curvature. In contrast to the GP method, deep learning exhibits enhanced capability in effectively reconstructing data beyond the observed range. Hence, we can make use of the full SGL systems and improve the precision of the constraints.

In this study, we developed a combined RNN and BNN architecture to accurately reconstruct the distance-redshift relation using the Pantheon sample. The RNN component of the network is specifically designed to predict the comoving distance at a given redshift, while the BNN component serves as a valuable complement, allowing for the calculation of uncertainties associated with these predictions. In the process of the distance reconstruction, we considered four activation functions and found that only tanh and selu functions can reproduce the Pantheon data well. Hence, we calibrated the distance of SGL systems with tanh and selu functions. To investigate the possible influence of different lens models on constraining the spatial curvature, we considered three types of lens models, i.e. the SIS model, PL model and EPL model. In SIS model, the spatial curvature is constrained to be $\Omega_k=0.049^{+0.147}_{-0.125}$ with tanh function, and $\Omega_k=0.082^{+0.125}_{-0.130}$ with selu function. Comparing with the result of \cite{Wang:2020}, $\Omega_k=0.57^{+0.20}_{-0.28}$, which favours an open Universe at $2\sigma$, our result favours a flat Universe with a higher accuracy due to the increase of available SGL data points. In PL model, a closed Universe is favoured, with the curvature parameter $\Omega_k=-0.245^{+0.075}_{-0.071}$ with tanh function, and $\Omega_k=-0.232^{+0.076}_{-0.068}$ with selu function, which is consistent with $\Omega_k=-0.246^{+0.078}_{-0.100}$ obtained in \cite{Wang:2020}. In EPL model, the spatial curvature is constrained to be $\Omega_k<0.011$ with tanh function, and $\Omega_k<0.051$ with selu function. Comparing with the results $\Omega_k=0.250^{+0.16}_{-0.23}$ in \cite{Wang:2020}, our constraint on the spatial curvature parameter is looser in EPL model, but there is no strong evidence ruling out a flat Universe. On the other hand, for the set of parameters in three lens models, the results demonstrate that the lens galaxies can not be simply described by the standard SIS model.

In summary, the lens mass models have noticeable influence on the curvature parameter. In SIS model, a spatially flat Universe is favoured within $1\sigma$ uncertainty. In PL model, a closed Universe is favoured at $\sim 3\sigma$ confidence level. In EPL model, constrain is relatively loose, but a flat Universe couldn't be excluded. More accurate modelling of the lens mass profile is necessary to further improve the constraint on the curvature parameter.

%
%
%

\bibliographystyle{raa}
\bibliography{bibtex}

\end{document}